# Shape Evolution of Sandwitched Droplet in Microconfined Shear Flow

Kaustav Chaudhury, Debabrata DasGupta, Tamal Roy and Suman Chakraborty[1]

*Department of Mechanical Engineering*

*Indian Institute of Technology*

*Kharagpur - 721302, INDIA*

[1]Corresponding author, e-mail: suman@mech.iitkgp.ernet.in

Droplets confined in a microfluidic channel often exhibit intriguing shapes, primarily attributable to complex hydrodynamic interactions over small scales. We show that effect of varied substrate wettability conditions may further complicate these interactions remarkably, and often non-trivially. Our studies reveal that the combined influence of substrate wettability and fluidic confinement eventually culminates towards influencing the droplet transients, distortion of the local shear flow field, as well as drop stabilization against breakup or detachment, allowing one to develop different regimes of shapes evolution that are fascinatingly distinct from the ones reported in earlier studies on drop breakup in micro-confined shear flows. We further demonstrate that combined consequences of wall effects and interfacial wettability characteristics can be exploited to pattern microfluidic substrates with pre-designed patches, bearing far-ranging scientific and technological consequences in several scientific and technological applications of contemporary relevance.

*Introduction* — From the purpose of optimizing artificially fabricated microfluidic systems to controlling blood flows in natural or artificial channels, it is critical to understand the dynamic behavior of deformable objects in confined geometries [1]. Morphological evolution of droplets in a continuous immiscible liquid phase in a microconfined flow offers with a representative picture that contains many intrinsic attributes of this generic, apparently simple yet elusively complicated physical scenario [2]. In spite of its vast scientific and technological relevance, however, the present state of understanding of confinement-induced dynamical evolution of droplets seems to be rather inadequate. This deficit stems from the complexities in understanding the combined consequences of the extent of confinement and substrate wettability on droplet morphodynamics, in response to an imposed flow condition. While researchers have examined droplet deformation in microconfined shear flows [2–7], no systematic studies have been reported in the literature on the effect of substrate wettability on such systems.

Fundamentally, droplet dynamics in an unconfined shear flow is well represented by Taylor's classical theory [8]. From a practical perspective, the same analysis applies when the droplet is located far-away from any solid boundary, reminiscent of an unconfined flow. However, when such conditions are not fulfilled, explicit effects of the walls need to be taken into consideration. Several recent studies, accordingly, have been reported in the literature to elucidate the hydrodynamic interaction between a drop in shear flow and its incipient confining boundaries [2–7]. Sibilo et al. [2] executed comprehensive studies for systematically investigating droplet breakup dynamics in a microconfined shear flow. However, investigation of any non-trivial implications of substrate wettability on the same remained beyond the scope of their study. Nevertheless, it can be appreciated that shape evolution of droplets in a confined micro-environment may not necessarily be dictated alone by the geometrical confinement in isolation, but is essentially governed by an intricate coupling between viscous stresses and interfacial tension as maneuvered by the substrate wettability-confinement coupling. This intricate coupling becomes highly relevant in microfluidic scenario, in which designed substrate wettability can be readily engineered by external physical or chemical treatment, or by external field effects (optical, electrical, thermal etc.).

Here, we report non-intuitive findings on shape evolutions of sandwitched liquid droplets in microconfined shear flow, under critical influence of substrate wettability variations. We capture the intricate coupling between substrate wettability and confinement-induced hydrodynamic interactions towards dictating the droplet morphodynamics. Our results effectively demonstrate that relative affinity of a droplet to the substrates defining the boundaries of the microfluidic confinement may alter the droplet pinchoff or detachment regimes in a rather profound manner. This, in turn, leads to a paradigm in which combinations of the extent of confinement and the substrate wettability can be precisely exploited to pattern the microchannel surfaces with desired patches originated out of remains from the pinched-off droplets, a paradigm that has hitherto remained unexplored.

*Models and Methods* — The present study on the dynamic evolution of a sanwitched droplet in a microconfined shear flow is carried out in an energy based mathematical framework of phase-field formalism where the distribution of the participating media (droplet and matrix in our case) are defined by the distributions of their relative phase concentrations, quantified by order parameter $\phi(\vec{r},t)$ over the entire domain. The Ginzburg-Landau free energy for a binary mixture can be expressed in terms of a phase-field parameter as [9–14] $F = \int_\Omega \left[ f(\phi) + \frac{K}{2}(\nabla \phi)^2 \right] d\vec{r}$, where $f(\phi) = \frac{A}{4}(\phi^2 - 1)^2$ is the bulk free energy density with $\phi(\vec{r},t) = \pm 1$ as the two stable solutions of the equilibrium order parameter profile corresponding to the matrix and the droplet domains respectively. The interfacial free energy coefficient ($K$) and the bulk free energy coefficient ($A$) can be related to interfacial thickness $\xi \sim \sqrt{K/A}$, and to the interfacial surface energy per unit area between the phases comprising the binary mixture $\sigma \sim \sqrt{KA}$ [9–14]. The dynamic evolution of the order parameter field is dictated by the Cahn-Hillard equation, which in the normalized form is given as [9–14]: $\partial_t \phi + \vec{u} \cdot \vec{\nabla} \phi = \nabla^2 \mu / Pe$. Here $\mu = \frac{1}{Cn}(\phi^3 - \phi) - Cn\nabla^2\phi$ is the driving chemical potential between the phases, such that $\mu = 0$ gives the equilibrium distribution of phases. The velocity field is divergence free and is obtained from the non-classical form of Navier-Stokes equation $\mathrm{Re}\, \rho_{eff}\left(\partial_t \vec{u} + u \cdot \vec{\nabla}\vec{u}\right) = -\vec{\nabla}p + \vec{\nabla} \cdot \left(\eta_{eff} \overline{\overline{D}}\right) + \left(\mu\vec{\nabla}\phi\right)/Ca$ where $\overline{\overline{D}}$ is the rate of deformation tensor $\left(\nabla\vec{u} + \nabla\vec{u}^T\right)$ and $\rho_{eff}$ & $\eta_{eff}$ are the effective phase averaged density and viscosity respectively. Here the additional body force $\mu\vec{\nabla}\phi/Ca$ is the non-classical phase-field term in Navier-Stokes equation representing hydrodynamic force of interaction (per unit volume) between the participating phases, realized over the diffused interface only, due to a interfacial jump in the generic $\nabla \phi$ term [9–14]. The important dimensionless groups are the Peclet number $Pe = \dot{\gamma} H^2 / M_C \sigma$, the Cahn number $Cn = \xi/H$, the Reynolds number $Re = \rho \dot{\gamma} H^2 / \eta$ and the capillary number $Ca = \eta \dot{\gamma} H / \sigma$ where $M_c$ is the critical mobility parameter for inter-diffusion of $\phi$, $H$ is the channel-height and $\dot{\gamma}$ is the shear-rate. The choice of these dimensionless groups is critical. The Peclet number, which denotes the relative importance of advection of $\phi(\vec{r},t)$ over diffusion, is set to $O(10^4)$ for the advection dominated phenomenon investigated here. Further, $Re$ is of the order of $10^{-1}$, and $Cn$=0.02. While we have studied results with numerous other choices of these parameters, we consider these values as mere representative parameters in order to bring out the essential new physics that we intend to highlight.

In order to bring out the explicit implications of substrate wettability in our simulation paradigm, we further note that the same can be ascertained by setting $\phi = \phi_s$ on the solid surface without alteration of any other governing equations and boundary conditions, where $\phi_s = \mp 1$ implies surface affinity towards the drop and matrix phases respectively and $\phi_s \in (-1,+1)$ implies partial affinity towards both the fluids. With this condition, the interfacial free energies of the phase transition points in terms of $\phi_s$ can be redefined to give an alternative definition of Young's equation in terms of local value of order parameter at the solid surface $(\phi_s)$ [11,14]: $\cos \theta_s = (\varphi_s^3 - 3\varphi_s)/2$.

In order to study the combined influence of surface wettability and fluidic confinement on dynamics of a sandwitched droplet, we perform numerical experimentations by solving the coupled Cahn-Hillard/Navier-Stokes equations discussed as above. The present interrogation domain consists of two parallel planes in two-dimensional space separated by a distance $H$. A droplet of diameter $d$ is sandwiched in between these two planes so that the condition: $d/H \geq 1$ is maintained. The droplet is allowed to relax so that it attains its equilibrium shape under quiescent condition, which is then used as the initial condition for the present study. Linear shear flow within the domain is imposed by translating both the planes in anti-parallel direction to each other with equal speeds so that a constant shear rate $\dot{\gamma}$ is maintained throughout. In the present analysis we keep viscosities $(\eta)$ and densities $(\rho)$ of both the droplet and matrix fluid same so that the effects of their contrast do not provide any additional artifacts.

*Shape evolution* — A droplet, when sandwitched in a microconfined flow, exhibits shape evolutions that are distinctively different from the morphological evolutions of non-sandwitched droplets under identical conditions. In case of sandwitched droplet, the surface-affinity conditions start playing significant role, often the lead role, towards dictating its dynamical evolution. The consequent interplay of wettability and confinement-induced hydrodynamic interactions gives rise to remarkably unique characteristics of droplet *detachment* and *pinchoff*, which have hitherto remained obscure.

In order to bring out features concerning the above in perspective, we first note that the presence of the solid surface alters the free energy in its vicinity. Depending on the free energy excess of the solid-liquid and solid-matrix interfaces, *detachment* or *pinchoff phenomena* are energetically favored. Forces acting on the droplet are: the viscous forces exerted by the surrounding fluid (attempting to deform the droplet), the surface tension forces (attempting to maintain the shape of the droplet), and the adhesive forces arising due to the interaction of the droplet/solid interface (attempting to drag the droplet along with the moving wall). Alteration of the surface affinity condition changes the surface adhesive forces and we observe two distinctive regimes of droplet dynamics (FIG 1). When the surface adhesive forces are not strong enough (typical to $\theta_S > 90°$), the surface tension force (which increases with increase in the droplet-matrix interfacial area) tends to dominate. Beyond a critical detachment time, the surface adhesive forces are not strong enough to hold the droplet along the surface, and detachment of the droplet takes place. The morphology of the droplet at the point of detachment (see FIG 1A(c), 1B) is best represented mathematically as a combination of double elliptic sigmoidal functions with tangents at the point of detachment given by the advancing and the receding angles of the contact line at that instant. On the other hand, with increase in the surface affinity (typical to $\theta_S < 90°$), the surface adhesive forces dominate the force competition and after a critical pinchoff time, the droplet pinchoff takes place in an attempt to minimize its surface energy. The droplet morphology in this regime (see FIG 1A(a), 1B) is more like a liquid bridge where the surface adhesive forces tend to hold the two ends of the bridge along the surface. Depending on the shear velocity and the velocity of the moving contact lines (which in turn depends on the surface affinity condition), the bridge elongates and pinches off, transferring part of the droplet to the surfaces and giving birth to a pinched daughter droplet.

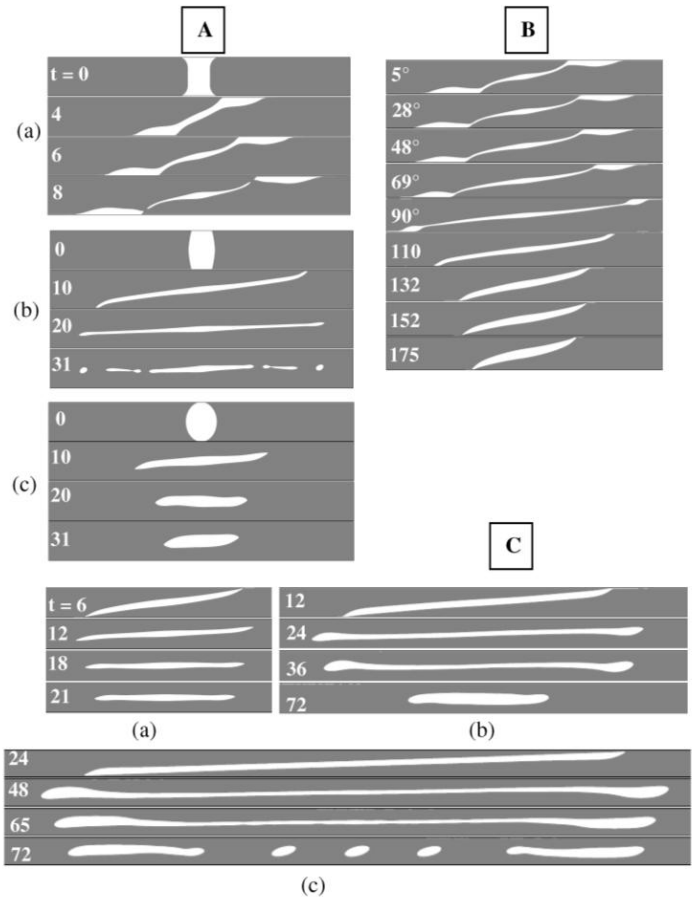

FIG 1. Typical characteristic features observed in the present study. (A) Typical representative behaviors (a) wall induced pinchoff (b) detachment and post detachment breakup (c) detachment and post detachment transience; (B) Effect of wettability variation: transition from pinchoff to detachment with decreasing surface affinity; (C) Effect of confinement and post detachment (in)stability for (a) d/H=1 (b) d/H=1.5 (c) d/H=2

The effect of confinement on the above-mentioned phenomena can be explained purely from hydrodynamic consideration. Greater confinement imposes a higher level of vorticity in the flow structure, for the same shear velocity. Presence of droplet in such a flow field causes a distortion in the streamlines and thereby produces nearly closed streamlines [see Ref. [2] and the references therein]. The shape and size of droplet can have a profound influence on distorting and closing the streamlines. Because of the high vorticity and closeness of the streamlines, the droplet dynamics is dominated by the Jaumann derivative component $\left(-\bar{\bar{\Omega}}\cdot\bar{\bar{S}}+\bar{\bar{S}}\cdot\bar{\bar{\Omega}}\right)$ rotating with the vorticity, rather than by the deformation component $\left(\bar{\bar{E}}\cdot\bar{\bar{S}}+\bar{\bar{S}}\cdot\bar{\bar{E}}\right)$, where $\bar{\bar{S}}$, $\bar{\bar{\Omega}}$ and $\bar{\bar{E}}$ are the droplet shape function [15], vorticity and deformation rate tensors. The degree of confinement by the streamlines, thus, dictates the evolution of the droplet (either pinched-off or detached) in a profound manner. As a consequence, we observe highly elongated shape evolutions in confined flows, which would otherwise be unstable in unbounded case [2].

It is important to mention in this context that because of a complicated interplay of confinement and wettability, dynamical evolution of the droplet in the present scenario cannot be explained in terms of the classical Taylor deformation parameter [8]. To circumvent this problem, we propose here a novel strategy for unique quantification of droplet morphological behavior based on droplet-matrix interfacial area strain $\left(\Gamma/\Gamma_0 - 1\right)$, where $\Gamma$ is the droplet-matrix interfacial area at instant *t* (time normalized with $\dot{\gamma}$) and $\Gamma_0 = \Gamma|_{t=0}$.

*Detachment and post-detachment behavior:* While exhibiting complex shape transiences within confined streamlines, a sandwiched droplet, in the detachment regime, can experience post-detachment breakup. While a droplet experiences breakup due to stretching in unbounded shear flow, breakup of a highly confined droplet $(d/H \geq 1)$ occurs due to drop shape instability; stretching is inhibited by close and confined streamlines [2–7]. To put the above in a quantitative perspective, in FIG 2 we first plot the droplet-matrix interfacial area strain at the point of detachment for different Ca, wettability and confinement. In FIG 2, we indicate the occurrence of post detachment breakup by filled markers, whereas we indicate the absence of such breakup by hollow markers. It is evident that at very high Ca (0.8 in the present study), the capillary dominance overwhelms and the post detachment-breakup is inevitable. However, at very low Ca (0.1 and 0.2 in the present study), the extent of confinement studied here is not sufficient to produce the stretching necessary for the post-detachment breakup. At intermediate ranges of Ca, the phenomenon is significantly more non-trivial. For example, at Ca=0.6, when the droplet is intuitively expected to suffer post detachment breakup, we observe stable droplet shape for d/H=1 and $\theta_s \geq 175°$. On the other hand, for higher levels of confinement (for example,

d/H=2), the confinement-induced stretching is sufficient enough to cause post-detachment breakup for the entire range of wettability investigated.

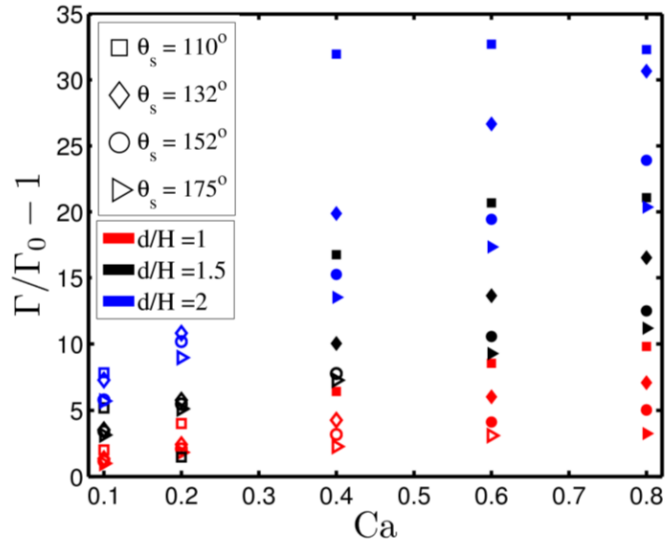

FIG. 2. Droplet-matrix interfacial area strain on detachment. Filled markers signify the breakup condition while the hollow markers represent no breakup situation.

From theoretical perspective, drop shape transience and breakup phenomena are an outcome of interplay of various shape instabilities [16,17]. Such instabilities have previously been investigated on the basis of idealized droplet geometries [see Ref. [18] and the references therein], which lose their pertinences in presence of the complex wettability-confinement coupling as addressed here. Under the present conditions, one may alternatively rely on the Fourier analysis [15] of the evolving drop shape function. Following the arguments of instability based theories [16,17], without any loss of generality, it can be mentioned that not all disturbances are allowed to grow as they are constrained by the interfacial area minimization. Strictly speaking, disturbances of very small frequencies, lying in between 0 to 1 (i.e., disturbances of long wavelength), can minimize the interfacial area and therefore are allowed to grow and finally lead towards droplet breakup. By observing the Fourier spectrums (see FIG. S1 in Supplementary Information at [URL]) we can see that near the breakup condition the dominant modes are lying in and around the frequency of unity. Thus, the stability criterion can be delineated by conditions where the dominating modes are shifted towards the high frequency regimes. A critical assessment, comparing FIGS. S1A (ii) and (iii) (See Supplementary

Information at [URL]), reveals that with increase in confinement effect, the dominant modes are shifted towards regimes having frequencies greater than unity. Such a shift supports the notion of 'confinement induced stabilization against breakup'. However, with an increase in confinement effect, dominating modes begin to grow in regimes characterized by frequencies lesser than unity (long wavelength disturbance situation). A situation may eventually occur in which such dominance results in successive breakup into multiple daughter droplets. Experimental evidences, indeed, suggest that under confined environment, the drop breakup occurs with higher elongation along with generation of more fragments [2–7]. The existence of long and short wavelength disturbances of comparable strengths (see FIG. S1A(iii) in Supplementary Information at [URL]), acting in tandem, demonstrates the theoretical basis of such observed behavior. Interestingly, surface affinity produces similar effects as that induced by the confinement effects (see FIG. S1B in Supplementary Information at [URL]), hallmarked by a dominating frequency shift towards higher frequency regime with an increasing surface affinity. Following this argument, the unusual stability of the droplet against breakup at Ca=0.6 and $\theta_S \geq 175°$ can be well appreciated, for d/H=1. Thus, it is not the confinement alone but the combined consequences of confinement and wettability that dictates the post-detachment morphological evolution characteristics of sandwitched droplets in a confined flow, which is a critical modification to the solely confinement dominated picture portrayed in previous studies [2–7].

*Spreading and pinchoff behavior-* The droplet dynamics in the spreading and pinchoff regime (typical to $\theta_S < 90°$) is analogous to that of liquid bridges with moving contact lines [19,20]. The surface adhesive forces tend to hold the two ends of the bridge along the surface. Effectively, the bridge is stretched depending on the shear velocity and the velocity of the moving contact lines. This, in turn, facilitates formation and growth of ligaments. During the elongation of the ligament, a neck region starts to form and keeps on collapsing. This eventually leads to pinchoff, transferring part of the droplet to the surfaces and giving birth to a pinched daughter droplet (inset to Fig. 3A schematically explains the underlying mechanism).

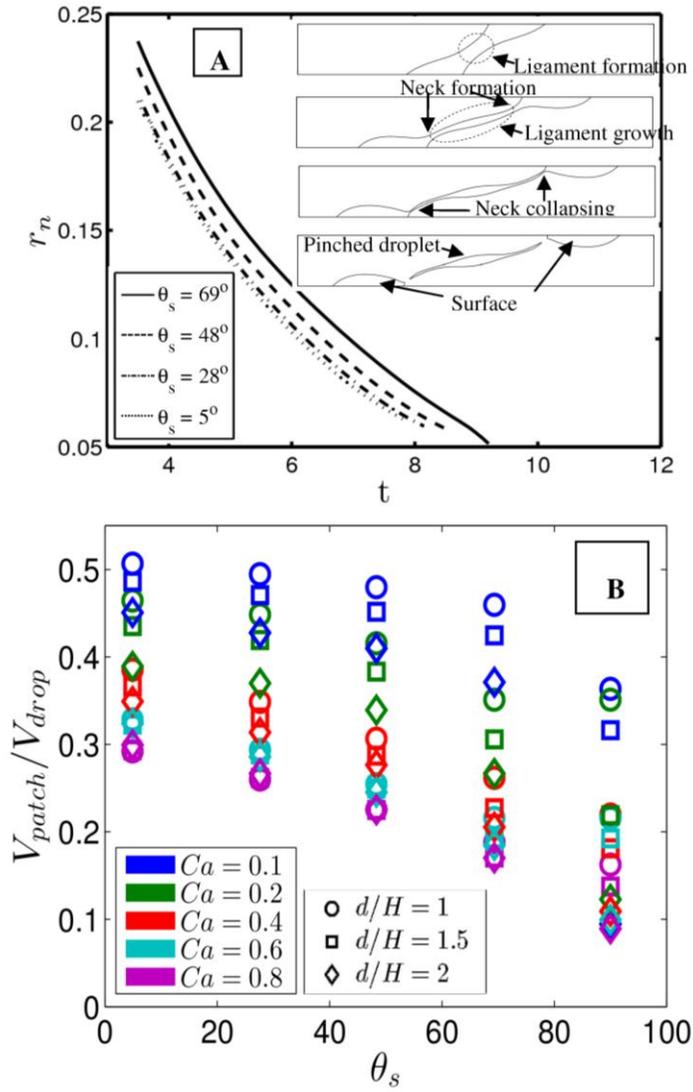

FIG. 3. Droplet morphodynamics in spreading and pinch-off regime under the combined effects of wettability and confinement in linear shear flow. In Fig. **A** we show the temporal evolution of neck radius ($r_n$) at Ca=0.4 and d/H=1. The inset shows the details of the spreading and pinch-off mechanism. Fig. **B** shows the variation of volume fraction of surface patches with surface wettability, confinement and Ca.

The liquid bridge is symmetric about the channel centerline because of the symmetry of the problem about this central axis and the fact that the droplet is unable to distinguish between the two confining boundaries, resulting in equipartitioning of the patch-volumes between the plates, irrespective of the surface wettability condition, extent of confinement or Capillary number.

Depending on the surface wettability condition and the extent of confinement, the temporal evolution of pinchoff mechanism differs significantly. Fig. 3A shows the temporal evolution of the neck radius ($r_n$) for different wettability conditions (typical evolution pattern is shown for Ca=0.4 and d/H=1). Irrespective of the surface wettability condition or the degree of confinement, the neck-radius decreases monotonically, till it reaches the pinchoff condition. With the reduction of surface affinity towards the droplet, however, the extent of droplet spreading over the channel surfaces reduces. Concurrently, the contact line speed relative to the substrates increases. As a combined consequence, the critical pinchoff time increases with reduction in the surface wettability. With increase in confinement, on the other hand, the thickness of the liquid bridge increases and the bridge can withstand shear upto a longer time. Accordingly, the critical pinchoff time increases with increase in confinement. It is, therefore, the combined effect of wettability and confinement that dictates the pinchoff dynamics and thus the size and shape of pinched daughter droplet as well as the extent of the undetached substrate-adhering layer. Regarding the role of *Ca*, it may be mentioned that higher the *Ca*, higher is the capacity of viscous force to deform and carry the droplet along with the flow direction and thus lower is the extent of spreading over the channel surface. Fig. 3B shows the variation of volume fraction of pinched daughter droplet with respect to the volume of the original droplet ($V_{patch}/V_{drop}$), under the combined influences of surface wettability, extent of confinement and *Ca*. With increase in confinement, the volume fraction of the surface-adhering patch decreases. On the other hand, with increases in *Ca*, the volume fraction of the patch decreases. With a judicious combination of confinement and wettability, in conjunction with *Ca*, one may, therefore, control the morphological evolution of undetached fluid patches on microfluidic substrates, in a pre-designed paradigm.

*Conclusions*- In summary, the present study attempts to bring out several non-intuitive features concerning shape evolutions of sandwitched droplets in a microconfined fluidic environment, under the combined consequences of the intrinsic substrate wettability as well as confinement-induced hydrodynamic interactions. This, in turn, leads to a new paradigm of controlling morphological evolutions of droplets in microfluidic environments. In the pinchoff regime, one may exploit the observed phenomena to pattern microchannel surfaces with pre-designed fluid patches contributed by the remains from the partially segregated droplets, through a combined influence of confinement and interfacial wettability; a paradigm that has hitherto remained unexplored.

# References

[1]     H. A. Stone, A. D. Stroock, and A. Ajdari, Annual Review of Fluid Mechanics **36**, 381 (2004).

[2]     V. Sibillo, G. Pasquariello, M. Simeone, V. Cristini, and S. Guido, Physical Review Letters **97**, 1 (2006).

[3]     M. A. Khan and Y. Wang, Physics of Fluids **22**, 123301 (2010).

[4]     M. SHAPIRA and S. HABER, International Journal of Multiphase Flow **16**, 305 (1990).

[5]     A. Vananroye, P. J. A. Janssen, P. D. Anderson, P. Van Puyvelde, and P. Moldenaers, Physics of Fluids **20**, 013101 (2008).

[6]     Y. Renardy, Rheologica Acta **46**, 521 (2006).

[7]     P. J. A. Janssen and P. D. Anderson, Physics of Fluids **19**, 043602 (2007).

[8]     G. I. Taylor, Proceedings of the Royal Society of London. Series A, Containing Papers of a Mathematical and Physical Character. **146**, 501 (1934).

[9]     V. E. Badalassi, H. D. Ceniceros, and S. Banerjee, Journal of Computational Physics **190**, 371 (2003).

[10]    R. Borcia and M. Bestehorn, Physical Review E **67**, (2003).

[11]    S. Chakraborty, Physical Review Letters **99**, 1 (2007).

[12]    R. Borcia and M. Bestehorn, Physical Review E **75**, 1 (2007).

[13]    H. Ding, M. N. H. Gilani, and P. D. M. Spelt, Journal of Fluid Mechanics **644**, 217 (2010).

[14]    R. Borcia, I. Borcia, and M. Bestehorn, Physical Review E **78**, 1 (2008).

[15]    The morphology of the droplet at the point of detachment is best represented mathematically as a combination of double elliptic sigmoidal functions with tangents at the point of detachment given by the advancing and the receding angles of the contact line at that instant.

[16] H. A. Stone and L. G. Leal, Journal of Fluid Mechanics **198**, 399 (2006).

[17] S. Tomotika, Proceedings of the Royal Society A: Mathematical, Physical and Engineering Sciences **150**, 322 (1935).

[18] M. Minale, Rheologica Acta **49**, 789 (2010).

[19] S. Dodds, M. D. S. Carvalho, and S. Kumar, Physics of Fluids **21**, 092103 (2009).

[20] S. Dodds, M. S. Carvalho, and S. Kumar, Journal of Fluid Mechanics 1 (2012).

# Supplementary Information: Shape Evolution of Sandwitched Droplet in Microconfined Shear Flow

Kaustav Chaudhury, Debabrata DasGupta, Tamal Roy and Suman Chakraborty[1]

*Department of Mechanical Engineering*

*Indian Institute of Technology*

*Kharagpur - 721302, INDIA*

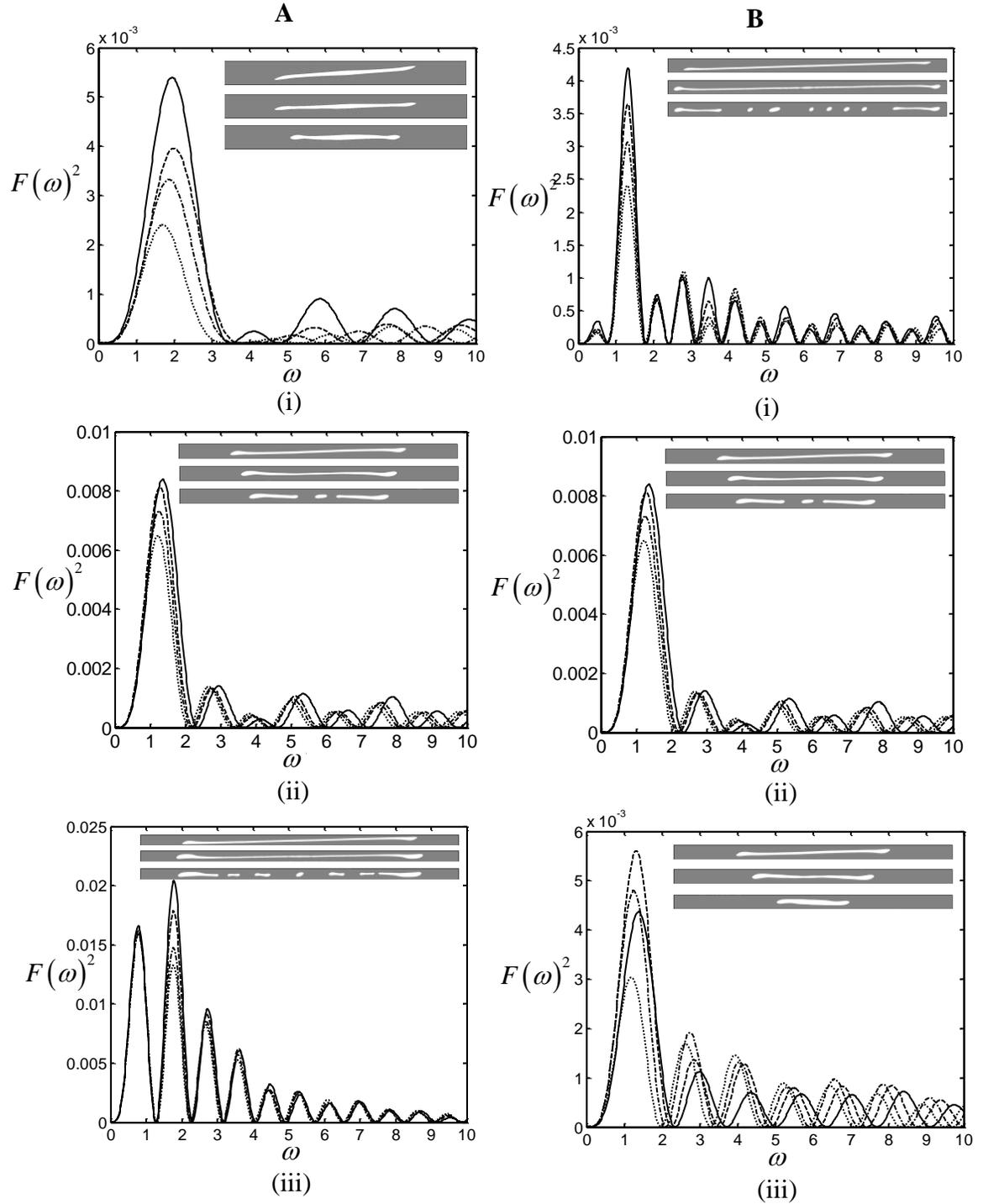

FIG. S1. Evolution of Fourier-spectrum of disturbances over the droplet-matrix interface upto the point of first pinchoff. The corresponding insets show some typical morphologies before and after pinchoff. The sequence of linestyles from dotted to chaindot to dashed to firm respectively represent increasing time. **A**. Effect of the variation of confinement (d/H): (i) 1, (ii) 2 and (iii) 3 for Ca = 0.4 and $\theta_s$ = 132°. **B**. Effect of the variation of wettability ($\theta_s$): (i) 110°, (ii) 132° and (iii) 152° for Ca = 0.4 and d/H= 1.5.